# Does the subtropical jet catalyze the mid-latitude atmospheric regimes?

Paolo M. Ruti[(1)], Valerio Lucarini[(2)], Alessandro dell'Aquila[(1)], Sandro Calmanti[(1)], Antonio Speranza[(2)],

[1] Climate Section, ENEA, Roma, Italy

[2] Dept. of Mathematics and Computer Science, University of Camerino, Camerino, Italy

## Abstract

Understanding the atmospheric low-frequency variability is of crucial importance in fields such as climate studies, climate change detection, and extended-range weather forecast. The Northern Hemisphere climate features the planetary waves as a relevant ingredient of the atmospheric variability. Several observations and theoretical arguments seem to support the idea that winter planetary waves indicator obey a non-gaussian statistics and may present a multimodal probability density function, thus characterizing the low-frequency portion of the climate system. We show that the upper tropospheric jet strength is a critical parameter in determining whether the planetary waves indicator exhibits a uni- or bimodal behavior, and we determine the relevant threshold value of the jet. These results are obtained by considering the data of the NCEP-NCAR and ECMWF reanalyses for the overlapping period. Our results agree with the non-linear orographic theory, which explains the statistical non-normality of the low-frequency variability of the atmosphere and its possible bimodality.

**1. Introduction**

The notion that well-defined winter mid-latitude atmospheric *patterns of flow* are *recurrent* during the northern hemispheric winters [*Dole, 1983*] has been repeatedly put forward, investigated and debated since the early definition of Grosswetterlage [*Baur, 1951*], to the classical identification of Atlantic blocking [*Rex, 1950*], all the way to more recent work on regimes detection and identification [*Corti et al., 1999*].

In such a perspective, the relevant problem concerning the general circulation of the atmosphere has been understanding whether the large scale atmospheric circulation undergoes fluctuations around a single equilibrium [*Nitsche et al., 1994; Stephenson et al., 2004*] or multiple equilibria [*Charney and Devore, 1979; Hansen and Sutera, 1986; Mo and Ghil, 1988; Benzi and Speranza, 1989; among others*]. The understanding of this climatic property would also help in responding to practical needs such as addressing the feasibility of extended range weather forecasts or the robust detection of climate changes [*Corti et al., 1999*].

Coming to the dominant physical processes, the mid-latitude dynamics feature upper tropospheric westerlies and synoptic to planetary waves as typical ingredients. The radiative forcing and the Earth rotation constrain the characteristics of the mean axially symmetric circulation and thereby the strength of the midlatitude westerly winds (jet) [*Held and Hou, 1980*]. The observational evidence shows that the strength of the westerlies is Gaussian-distributed ($u_{mean} \approx 30 m\ s^{-1}$). Instead, it has been proposed that the activity of the ultra-long planetary waves may have a bimodal distribution [*Hansen and Sutera, 1986*], although the issue of the statistical significance of non-unimodal distributions of planetary-wave amplitudes has been intensely debated [*Nitsche et al. 1994; Hansen and Sutera 1995; Stephenson et al., 2004; Christiansen 2005*].

A theoretical support exists for the search of multimodal distribution of the activity of planetary waves. In the context of orographic resonance theories, the zonal flow – wave field interaction (via form-drag) was first proposed as a driving mechanisms allowing for the occurence of multiple equilibria of the planetary waves amplitude [*Charney and Devore, 1979*]. However, transitions between the quasi-stable equilibria require for energetic reasons large variations ($\Delta u \approx 40 m\,s^{-1}$) of the mean westerlies, at odds with the "normality" of the distribution of the observed westerlies strength [*Malguzzi and Speranza, 1981; Benzi et al., 1986*].

Therefore, a dynamical interpretation featuring fixed strength of the westerlies was required. Benzi et al. [1986] suggested that, when the mean westerlies are subresonant or superresonant, the response of mid-latitude atmospheric long waves to orographic forcing can be well described in terms of linear Rossby waves. Instead, near topographic resonance the meridional structure of the zonal wind may produce a wave self non-linearity leading to multiple equilibrium amplitudes of the perturbation field [*Malguzzi et al. 1997*]. Thus, the bent resonance curve not only explains the existence of the multiple equilibria of the planetary wave amplitude, but also suggests that relatively small changes of the jet strength may imply a switch from unimodal to multimodal regimes of the atmospheric circulation (see Figure 1).

## 2. Data and methods

In this study, we consider the two major reanalysis products released by NCEP-NCAR [*Kistler, et al, 2001*] and by ECMWF [*Simmons and Gibson, 2000*] (hereafter NCEP and ERA40, respectively). In particular, we consider the winter-time daily fields (DJF) of the 500 hPa geopotential height and of the 200 hPa zonal wind, for the overlapping time frame ranging from December 1st 1957 to August 31$^{st}$ 2002.

Two proper counterparts to the dynamical parameters employed in the theories are extracted from such datasets by computing two robust indicators of relevant large scale features of the mid-latitude troposphere. The Wave Activity Index (WAI), introduced by Hansen and Sutera [*1986*], is computed as the root mean square of the zonal wavenumbers 2 to 4 of the winter 500 hPa geopotential height variance over the channel 32°N – 72°N. The WAI provides a synthetic picture of the ultra long planetary waves and captures the orographic resonance, since an approximate mode of zero phase velocity (resonance) is 3 [*Malguzzi and Speranza, 1981*]. The Jet Strength Index (JSI) is computed daily as the maximum of the zonal mean of the zonal wind at 200hPa, where the sub-tropical jet peaks.

In order to filter out the synoptic atmospheric variability, we apply a low-pass filter to both WAI and JSI indexes by performing a 5-day running mean to the signal. In order to capture the anomalies with respect to the seasonal cycle we filter out from the WAI signal the most prominent spectral peaks occuring at 12, 6 and 4 months, which are directly related to influence of the external solar forcing. On the other hand, in the context of the bent resonance theory [*Benzi et al., 1986*] the jet strength is considered as an autonomous forcing parameter of the system, controlling and catalysing its internal variability. Therefore, we do not filter out from the JSI the seasonal cycle and its harmonics as done for the WAI. The signature of the synoptic atmospheric variability is removed by performing a 5-day running mean on both WAI and JSI indexes. Furthermore, the results are robust with respect to the different filtering techniques.

The continuous probability density functions have been computed using a kernel estimation technique depending on a smoothing parameter h [Silverman, 1986].

**3. Results.**

Slight discrepancies between the two reanalyses are expected in view of the different description of midlatitude wave activity discussed by Dell'Aquila et al. [2005]. Therefore, we first assess the overall equivalence of the picture provided by the winter WAI and JSI extracted from the two reanalyses, by performing one- and two-dimensional Kolmogorov-Smirnov test [*Fasano and Franceschini, 1987*] on the distribution of the single variables and on the joint PDFs. In both cases, the PDFs are equivalent at a confidence level larger than 95%. Therefore, we can safely consider the results obtained with NCEP as representative for both reanalyses and highlight the main differences where necessary.

The empirical joint PDF, constructed by means of two-dimensional gaussian estimators (Figure 2) presents multiple, well defined, peaks distributed over the WAI-JSI space. A major peak (point A in Figure 2) corresponds to weak upper tropospheric jet (JSI ~ 40 m s$^{-1}$) and low-to-intermediate activity of the planetary waves (WAI ~ 60 m). A second peak, (point B) corresponds to intermediate values of JSI and low values of WAI (JSI ~ 43 m s$^{-1}$; WAI ~ 55 m). The third peak (point C) corresponds to similar intensities of the jet (JSI ~ 45 m s$^{-1}$) and very high values of (WAI ~ 70 m). The fourth peak (point D), corresponding to intense sub-tropical jet (JSI ~ 50 m s$^{-1}$), features relatively weak activity of planetary waves (WAI ~ 55 m). These features are consistent with the theoretical framework sketched in Figure 1, where three different regions can be separated, characterized by low, intermediate and high intensities of the tropospheric jet and by a different number of equilibrium amplitudes of the planetary waves.

While the 2d joint PDF essentially provides a qualitative view on the properties of the system, a more stringent statistical interpretation of such analogy can be highlighted by considering the distribution of WAI obtained by fixing the range of JSI variability. We split the entire WAI-JSI space into three sectors, characterized by low (38 m s$^{-1}$ < JSI < 41 m s$^{-1}$), intermediate (42 m s$^{-1}$ < JSI < 47 m s$^{-1}$) and high (48 m s$^{-1}$ < JSI < 52 m s$^{-1}$) intensities of the jet, each sector comprising about 1/6, 1/3, and 1/6 of the total sample population, respectively. The empirical distributions of WAI (Figure 2b) are meant to be a statistically robust representation of the atmospheric planetary waves in each sub range of the JSI values. Well-distinct peaks are observed in the intermediate range, while for weak or strong JSI unimodal distributions appear. Notice that strong oceanic tropical forcing (El Nino) [*Philander, 1990*] implies zonal elongation and strengthening of the sub-tropical jet respect to normal conditions, resulting in strong JSI (more than 48 m/s), and locating the El Nino years in the upper region of the phase-space (Figure 2a).

Statistical robustness in the properties of the PDFs is required in order to avoid artificial results [*Stephenson et al., 2004*]. Since we are testing the hypothesis of having a specific number of peaks in each of the considered JSI sub-range, we estimate the optimal kernel width $h_0$ by generating an ensemble of surrogate datasets (1000 members) and then choosing the value of $h$ that maximizes the trade-off between having as many surrogate distributions with the correct number of peaks and as few surrogate distributions with the wrong number of peaks. The surrogate datasets are generated with a bootstrap Montecarlo experiment in which 45 winters are selected randomly with repetition among the reanalysis period lasting from 1958 to 2002 for both JSI and WAI at the same time. We stratify the data according to the values of JSI and, after spanning a whole range of values of $h$ we find that the best trade-off is realized for $h = h_0 = 3.5 m$ with a broad maximum ranging from $h = 3m$ to $h = 4m$, in close agreement with a recent study, where a different measure of the skill score was considered [*Christiansen, 2005*]. For this value of $h$ we have that among the surrogate datasets extracted from NCEP (ERA40), in the low-JSI range 94% (68%) of the realizations have unimodal distribution of WAI; in the intermediate-JSI range 88% (82%) of the realizations have bimodal distribution and in the high-JSI range, 79% (55%) of the realizations have unimodal distribution.

It is important to test null-hypothesis of unimodality for the intermediate range of JSI. A first test is performed by constructing a Gaussian distribution equivalent to the WAI pdf for intermediate JSI. A second test is performed by considering the unimodal distribution obtained by increasing the initial smoothing parameter in the WAI pdf until a unimodal distribution is attained. For both NCEP-ERA40 reanalysis, such marginal kernel width $h_u$ is 5 m. From these two unimodal distributions, we extract a set of 10000 WAI surrogate winter time series and then, computing each time the pdf with the kernel width $h_0$, we count the fraction of the computed WAI pdfs having a dip larger than that obtained with the original data. For the test with the Gaussian-equivalent distribution, we have that for both NCEP and ERA40 datasets, the bimodality is statistically significant at the 99.9% level. For the test with the marginal unimodal distribution, the null-hypothesis can be rejected with over 97% (91%) confidence level for the NCEP (ERA40) dataset. These results are only weakly sensitive to the choice of the kernel width. Since our 1D pdf results from selecting days with inter-mediate JSI values, it is not the result of a continuous time-series. So, it is intrinsically impossible to take into account time lagged correlation as done in Christiansen (2005) when assessing the significance of the 1-D pdf properties. Therefore, the statistical confidence of this test may be overestimated.

The position and height of the peaks of wave activity corresponding to each class of intensity of the jet are safely characterized in a statistical sense. In particular, the two peaks observed for intermediate values of JSI are well separated (~ 20 m). Moreover, the two reanalyses agree in the position of the peaks, despite the different degree of statistical confidence. Our results appear robust since similar and consistent results – although with different degree of statistical confidence – are obtained with all the values of $h$ ranging from 2 m to 5 m.

Using the density PDFs given in Figure 2b to stratify the data, the time-mean maps of the zonal anomalies for all days corresponding to the peaks of the bimodal distribution and of the unimodal distributions have been computed. For peaks A and D (Figure 3 i-iii), we select the days characterized by a range of WAI values associated to a population which is higher than half the height of the corresponding distribution. For peaks B and C, we consider the days falling in a range of WAI values that starts at the dip value of the bimodal distribution and extends to twice the distance from the associated relative maximum.

The difference map between the two intermediate JSI patterns B and C (Figure 3-ii) indicates a more amplified wave number three component for the pattern C, with higher geopotential height centers over the northern Pacific and Alaska, the Northern Sea, and the Siberian land. Previous analyses [*Hansen and Sutera, 1995; Christiansen 2005*], which did not stratify data with the jet strength, highlighted a predominance of the wave number two in the difference map. The eddy field corresponding to the low JSI (pattern A) shows the lowest ridge over the Rockies and the higher Greenland through, while for the high JSI (pattern D) the highest ridge over the Rockies is observed.

**4. Conclusions.**

Our analysis indicates that in the Northern Hemisphere the statistics of the planetary atmospheric waves can be characterized in terms of the sub-tropical jet, consistently with the physical framework proposing the nonlinear modification of the topographic resonance due to the nonlinear wave self-interaction [*Benzi et al., 1986*] as the basic mechanism for the low-frequency variability of the atmosphere. Our physically-based approach is somewhat complementary to the dynamical system-based approach presented in Christiansen [2005]. We have proved on the available NCEP and ERA 40 global reanalyses that for intermediate jet strength, an indicator of the planetary waves presents a bimodal behavior, while for higher or lower values of the jet strength the estimated pdf is unimodal.

Thus, the interpretation of the interaction between the tropics and the mid-latitudes should be addressed not only considering the wave trains emanating from the tropics [*Hoskins and Ambrizzi, 1993*], but also in the perspective of the role of the sub-tropical jet in defining the statistical properties of the planetary waves. The zonal mean circulation (Hadley circulation), and the tropical oceanic forcings (i.e. ENSO) play a relevant role in this view. In this perspective, the strongest El Nino years could be located in the upper branch of a hysteresis cycle, where the system undergoes a unique solution. Similar conclusion has been obtained by Molteni et al. [*2005*].

# References

Baur, F. (1951), Extended range weather forecasting. *Compendium of Meteorology*, Amer. Meteorol. Soc., 814-833.

Benzi, R., Speranza, A. (1989), Statistical properties of low frequency variability in the Northern Hemisphere. *J. Climate 2*, 367-379.

Benzi R., Malguzzi., P., Speranza, A., Sutera (1986), A. The statistical properties of general atmospheric circulation: observational evidence and a minimal theory of bimodality. *Quart. J. Roy. Met. Soc.*, *112*, 661-674.

Charney, J. G., Devore, J.C. (1979) Multiple flow equilibria in the atmosphere and blocking. *J. Atmos. Sci.*, *36*, 1205-1216

Christiansen, Bo, (2005) On the bimodality of planetary-scale atmospheric wave amplitude index. *J. Atmos. Sci.*,. In press.

Corti, S., Molteni F., Palmer, T.N. (1999) Signature of recent climate change in frequencies of natural atmospheric circulation regimes. *Nature*, *398*, 799-802.

Dell'Aquila, A., Lucarini, V., Ruti, P.M., Calmanti S. (2005) Hayashi Spectra of the Northern Hemisphere Mid-latitude Atmospheric Variability in the NCEP-NCAR and ECMWF Reanalyses. *Climate Dynamics* DOI: 10.1007/s00382-005-0048-x.

Dole, R. M. (1983) Persistent anomalies of the extratropical Northern Hemisphere wintertime circulation. *Large-Scale Dynamical Processes in the Atmosphere,* B. J. Hoskins and R. P. Pearce, eds., Academic Press, NY, 95-109.

Fasano, G., Franceschini (1987), A. A multidimensional version of the Kolmogorov-Smirnov test . *Mon. Not. R. Astr. Soc. 225,* 155–170.

Hansen, A.R., Sutera, A. (1986) On the probability density distribution of Planetary-Scale Atmospheric Wave amplitude. *J. Atmos. Sci., 43,* 3250-3265.

Hansen, A.R., Sutera, A. (1995) The probability density distribution of Planetary-Scale Atmospheric Wave amplitude Revisited. *J. Atmos. Sci., 52,* 2463-2472.

Held, I.M, Hou, A.Y. (1980) Nonlinear axially symmetric circulations in a nearly inviscid atmosphere. *J. Amos. Sc..., 37,* 515-533.

Hoskins, B.J., Ambrizzi, T. (1993) Rossby wave propagation on a realistic longitudinally varying flow. *J. Atmos. Sc.*, *50,* 1661-1671.

Kistler R, et al. (2001) The NCEP-NCAR 50-year reanalysis: Monthly means CD-ROM and documentation. *Bull. Am. Meteorol. Soc. 82,* 247–267

Malguzzi, P., Speranza, A. (1981) Local Multiple Equilibria and Regional Atmospheric Blocking. *J. Atmos. Sc.*, *9*, pp. 1939–1948.

Malguzzi, P., Speranza A., Sutera A., Caballero R. (1997) Nonlinear amplification of stationary Rossby waves near resonance, Part II. *J. Atmos. Sci.*, *54*, 2441-2451.

Mo, K., Ghil, M. (1988) Cluster analysis of multiple planetary flow regimes. *J. Geophys. Res.*, *93*, 10927-10952.

Molteni, F., F. Kucharski and S. Corti, (2005) On the predictability of flow-regime properties on interannual to interdecadal timescales. In: "Predictability of weather and climate", T.N. Palmer and R. Hagedorn Eds., *Cambridge University Press*. In press.

Nitsche, G., J. M. Wallace, and C. Kooperberg (1994) Is there evidence of multiple equilibria in planetary wave amplitude statistics? J. Atmos. Sci., *51*, 314–322.

Philander, S.G. (1990) *El Nino, La Nina, and the Southern Oscillation*. San Diego, Academic Press

Rex, D.F. (1950) Blocking action in the middle troposphere and its effect upon regional climate. Part 2: The climatology of blocking action. *Tellus*, **2**, 275-301.

Silverman, B.W. (1986) Density *Estimation for Statistics and Data Analysis*. Chapman & Hall.

Simmons, A. J., Gibson, J.K. (2000) The ERA-40 Project Plan, ERA-40 Project Report Series No. 1, ECMWF, 62 pp.

Stephenson, D.B., Hannachi, A., O'Neill, A. (2004) On the existence of multiple climate regimes. *Quart. J. Roy. Mel. Soc.*, *130*, 583-605.

**Acknowledgments.**
The authors wish to thank A. Sutera and two anonymous referees for useful suggestions. NCEP data have been provided by the NOAA-CIRES Climate Diagnostics Center (http://www.cdc.noaa.gov/). The ECMWF ERA-40 data have been obtained from the ECMWF data server.

**Figure 1**. Qualitative behavior of the amplitude of the orographic wave (WAI) as a function of an indicator of the jet strength (JSI). Stable equilibria are indicated by the large dots.

**Figure 2**. a) Two-dimensional joint pdf using the index of the planetary waves (WAI) and the index of the tropospheric jet strength (JSI). b) One-dimensional pdf of the planetary waves index (WAI) for low JSI (red), intermediate (blue) and for high JSI (black). NCEP dataset, 1958-2002 winters. Units: WAI [m], JSI [m/s].

**Figure 3**. 500 hPa geopotential height eddy composites corresponding to the peaks in Fig2a. Panel i) and iii) for the A and D peaks. In panel ii), the difference between the composites for B and C peaks. Units are m and contour interval is 50 m for i) and iii), and 10 m for panel ii). Dashed lines for negative values. The shading is for the related 200hPa zonal wind speed U (Units - m/s, contour interval 10 m/s).

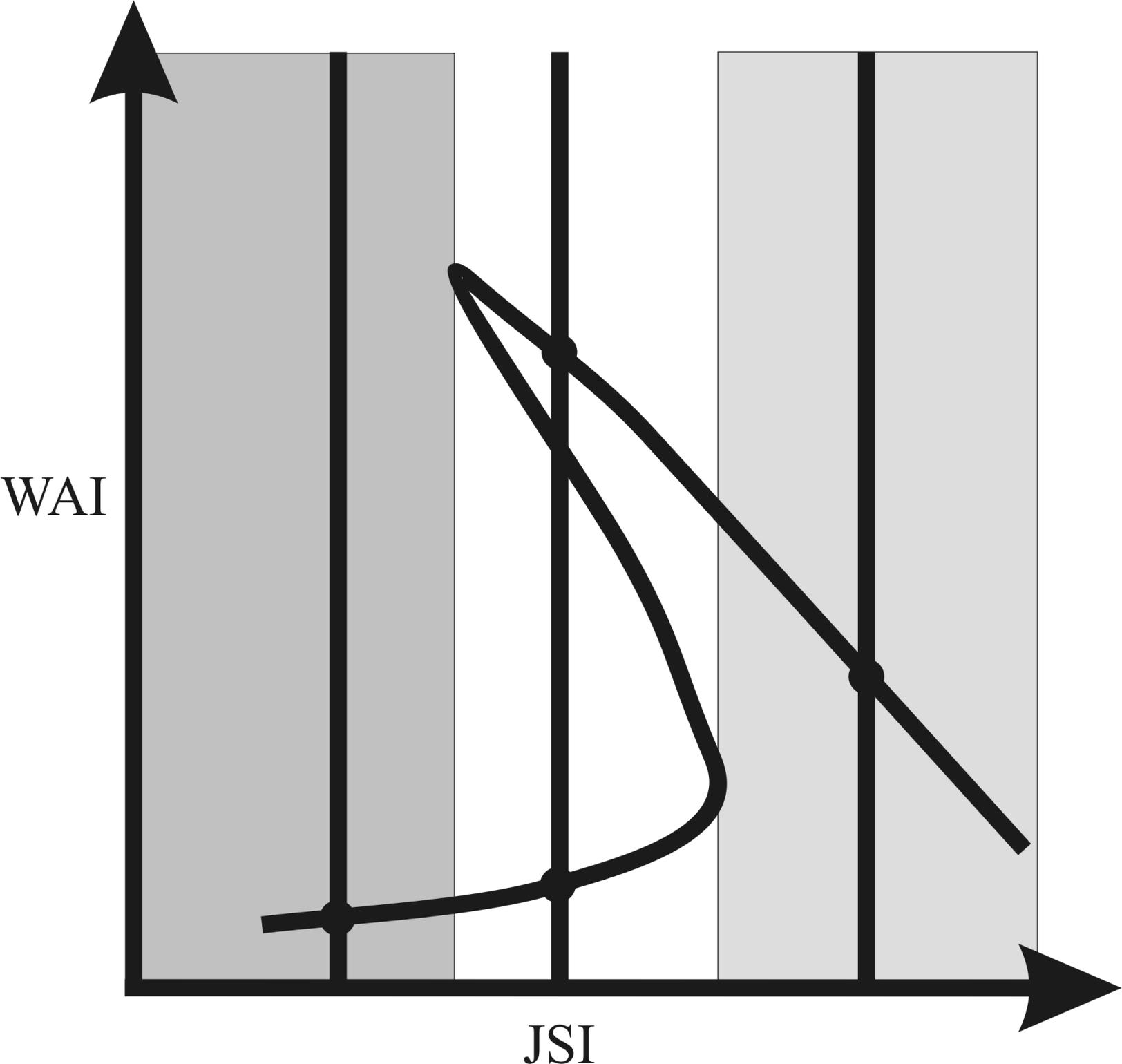

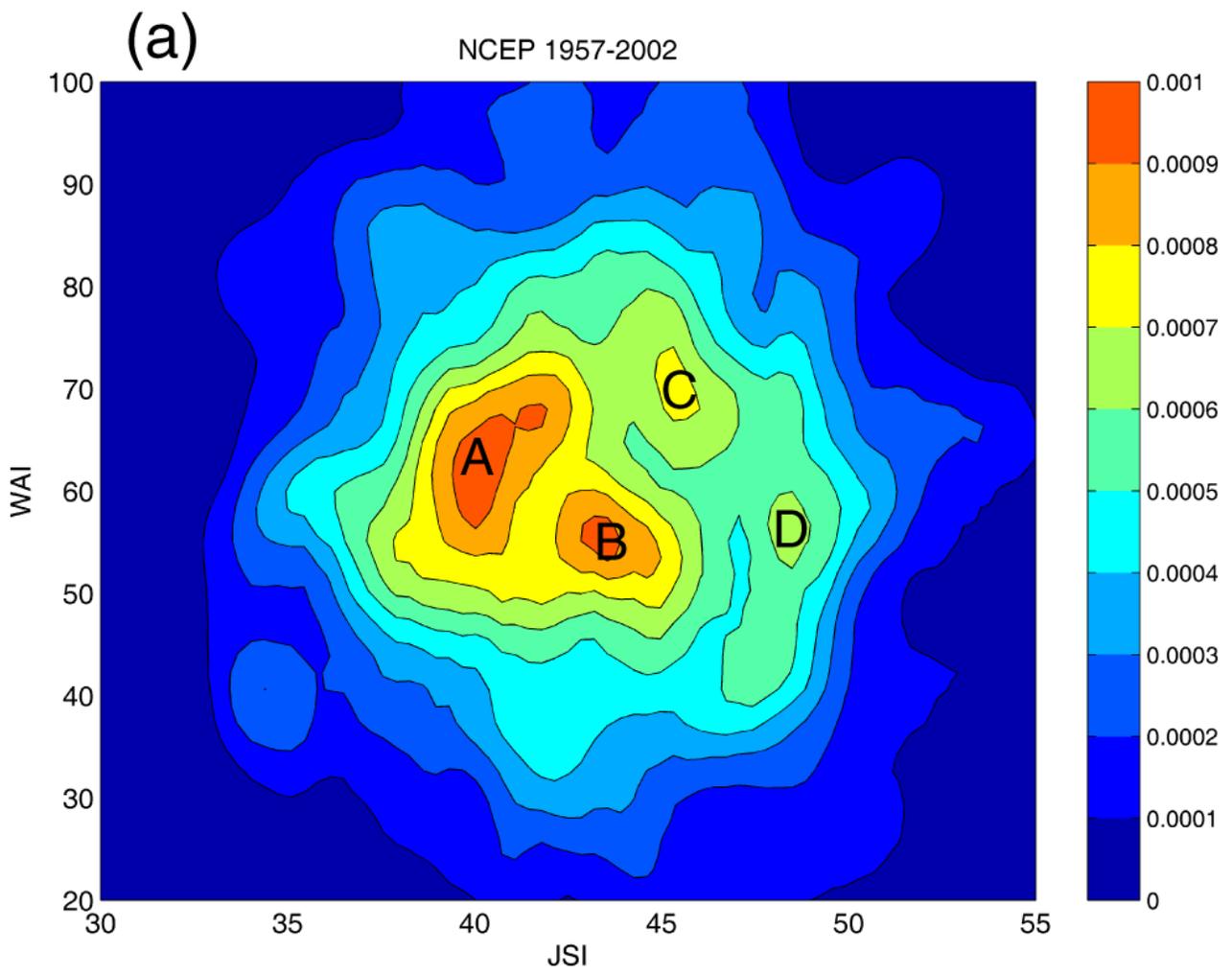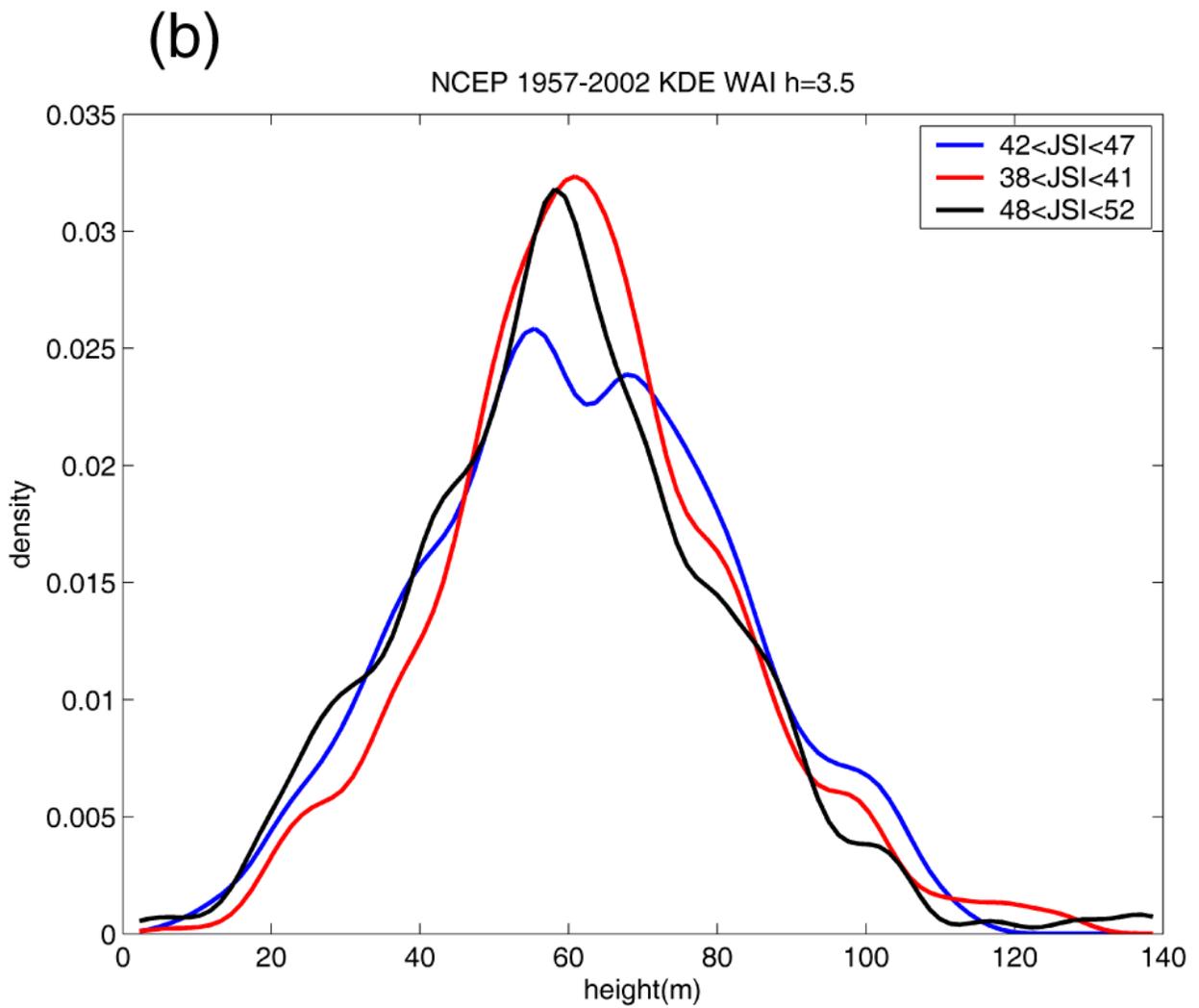

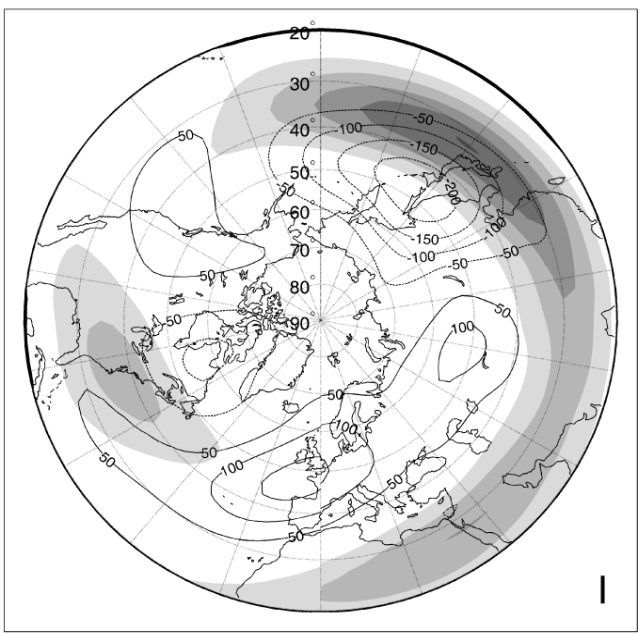
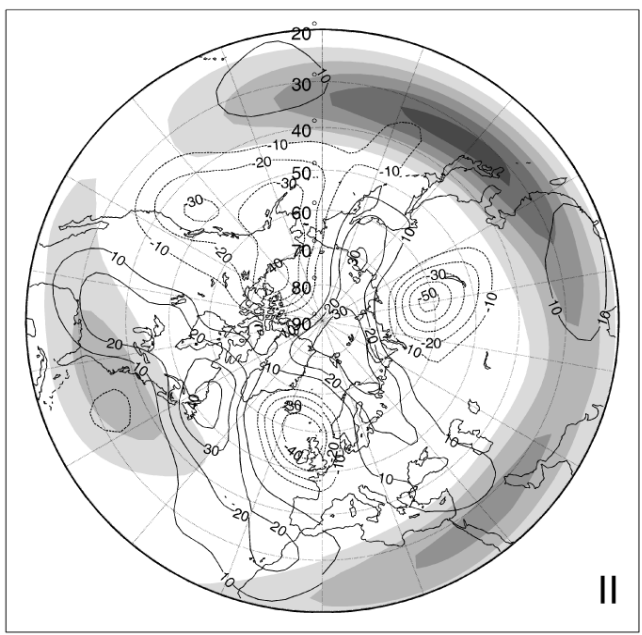
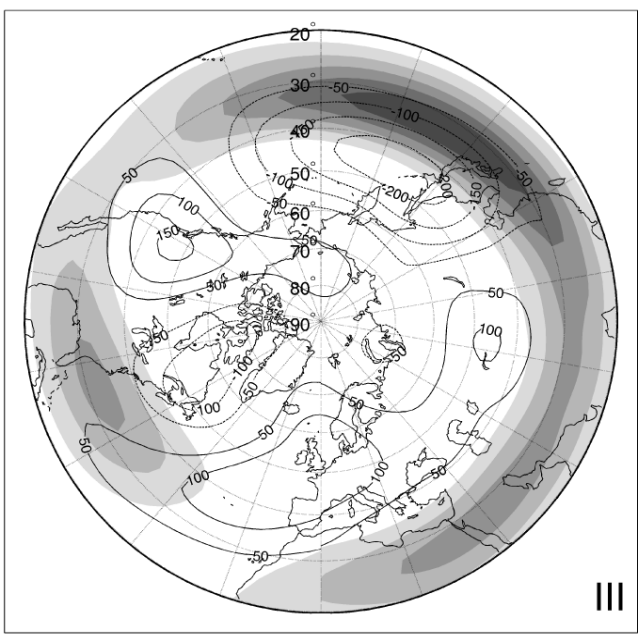
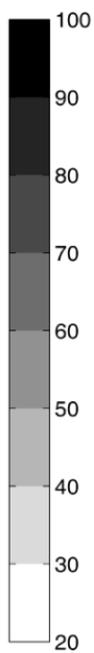